\documentclass[a4paper,11pt]{article}
\usepackage{pos}
\usepackage{graphicx}
\usepackage{gensymb}
\usepackage{booktabs}

\title{The H.E.S.S. Gravitational Wave and Gamma-Ray Burst Follow-Up Programs}

\author*[a]{Bernardo Cornejo}
\author[b, c]{Halim Ashkar}
\author[d]{Matteo Cerruti}
\author[a]{Ilja Jaroschewski}
\author[d]{Pierre Pichard}
\author[d]{Santiago Pita}
\author[a]{Fabian Schussler}

\onbehalf{on behalf of the H.E.S.S. Collaboration}

\affiliation[a]{IRFU, CEA, Université Paris-Saclay,\\
  F-91191, Gif-Sur-Yvette, France}

\affiliation[b]{Institute of Space Sciences (IEEC-CSIC),\\
Campus UAB, Torre C5, 2a planta, 08193 Barcelona, Spain}

\affiliation[c]{Laboratoire Leprince-Ringuet, École Polytechnique (UMR 7638, CNRS/IN2P3, Institut Polytechnique de Paris),\\
91128 Palaiseau, France}

\affiliation[d]{Université Paris Cité, CNRS, Astroparticule et Cosmologie,\\
F-75013 Paris, France}

\emailAdd{bernardo.cornejo@cea.fr}
\emailAdd{hess-grb-team@mpi-hd.mpg.de}

\abstract{Multi-wavelength and multi-messenger astrophysics have experienced rapid growth over the past decade, seeking a complete picture of different cosmic phenomena. Transient sources, in particular, benefit from the input of multi-messenger observations, offering complementary perspectives on the same event while maximizing the detection probability of a rapidly fading signal.
In this context, Gravitational Wave (GW) detections serve as perfect triggers for potential counterpart detections. Notably, a GW alert could be associated with a Gamma-Ray Burst (GRB), jetted cataclysmic events produced either by the collision of a binary neutron star system or a core-collapse supernova. These sources also radiate across the electromagnetic spectrum, allowing detection by X- and gamma-ray instruments aboard various satellites and thus enabling multi-wavelength triggering opportunities. The strong interest in minimizing reaction time to capture the full-time evolution of the emission, together with the often challenging localization uncertainties of the alerts, underscores the need for rapid and well-coordinated follow-up programs such as the one developed by the H.E.S.S. Collaboration.

This contribution will give an overview of the transient follow-up strategy carried out by the H.E.S.S. Collaboration, from the external alert trigger and the automatic reaction of the observatory to the various analysis steps of the obtained observations. To illustrate this comprehensive strategy, we will show two examples of follow-up observations of both GRBs and GWs, highlighting key results and challenges in the search for an associated high-energy gamma-ray emission.}

\ConferenceLogo{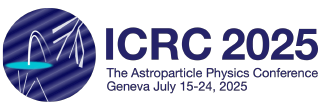}

\FullConference{39th International Cosmic Ray Conference (ICRC2025)\\
 15–24 July 2025\\
Geneva, Switzerland\\}

\begin{document}
\maketitle

\section{Introduction}

\subsection{The H.E.S.S. High Energy Gamma-Ray Observatory}

The High Energy Spectroscopic System (H.E.S.S.) is an imaging atmospheric Cherenkov telescope array (IACT) located at an elevation of 1800 m above sea level on the Khomas Highland plateau of Namibia (23$^\circ$16'18'' South, 16$^\circ$30'00'' East). The original array, operational since 2004, consists of four telescopes (CT1-4), each equipped with a 12-meter diameter segmented mirror. These telescopes operate in stereoscopic mode in order to achieve sensitivity to very-high-energy (VHE) gamma rays in the range of 100 GeV to 10 TeV. In 2012, a fifth telescope (CT5), with a 28-meter diameter segmented mirror, was added to the array. It was conceived to extend the accessible energy range towards lower energies and to allow for rapid slewing. For the study presented here, we only use data from CT1–4, analyzed in stereoscopic mode.  

\subsection{Motivation}

Gamma-Ray Bursts (GRBs) are extremely energetic explosions, capable of releasing up to $10^{55}$ erg in a few seconds. Bursts can last up to two seconds for short GRBs (sGRB) or until several minutes for long GRBs (lGRB), usually followed by a longer "afterglow" from radio to VHE $\gamma$-rays. Although mechanisms at the origin of this phenomenon are still under discussion, it is believed that lGRB are the product of massive stellar core-collapse supernovae, while sGRB are the consequence of compact binary systems merging (involving at least one neutron star). The emission from both is currently explained by the Fireball-Model \cite{fireball}, which argues that the newly formed compact object acts as a central engine powering an ultra-relativistic jet. Internal shocks within the jets would produce shock-accelerated material responsible for the prompt emission, while the afterglow would be generated as the jet interacts with the interstellar medium. The object can become opaque to $\gamma$-rays if the surrounding photon field density is high enough to allow for $\gamma$-$\gamma$ absorption. Moreover, at high redshifts, VHE photons will be significantly absorbed through interactions with the Extragalactic Background Light (EBL). VHE photons from GRBs are thus elusive, and only a few objects have been detected in energies above 100 GeV. This was the case for H.E.S.S. with GRB 180720B \cite{grb_hess_1} and GRB 190829A \cite{grb_hess_2}. These few events allow to place constraints on emission models and have shown puzzling characteristics, like how the VHE gamma-ray spectrum seems to decrease at the same rate as the one from X-rays.

Detection of Gravitational Waves (GWs) emitted by binary coalescences, such as the ones seemingly at the origin of sGRB, also provides crucial information about its precursor. Masses of the two merging objects, inclination of the system before merging, involved spins, and distance of the merger are all information provided by the detected GW signal. If the merger involves a neutron star, a GW detection can also provide us with information on the energy going into the ejecta. If VHE gamma-rays are detected, they will provide key insight in particle acceleration processes in jets/shocks. A complete understanding of the most energetic phenomena in the Universe therefore calls for a joint study of the different 'cosmic messengers' through multi-messenger and multi-wavelength observations.

The temporal and spatial challenges (coming from the often large localization uncertainties) presented by these types of alerts require a fast and accurate follow-up strategy in order to maximize the probability of a prompt VHE detection, such as the one developed by H.E.S.S. over the years.

\section{The H.E.S.S. Transient Follow-Up Program}

Follow-up observations of GRBs and GWs with H.E.S.S. are managed by a fully automated pipeline integrated into the central data acquisition system (DAQ) \cite{hess_daq}. This Target of Opportunity (ToO) alert system \cite{hess_follow_up_system} processes source information received from public alerts distributed via the General Coordinates Network (GCN\footnote{https://gcn.nasa.gov/}) circulars and notices. The alert passes through event-specific filters defined by each working group according to the scientific goals set by the collaboration. If the event meets the observation criteria, a notification is sent to the shifters and members of the working group. If the event is visible at the time of the alert, the system automatically triggers a response, forwarding follow-up instructions to the central telescope steering software, which is capable of overriding the current observation schedule to target the event location. If it is not observable on the same night, a revised strategy is discussed within the working groups, and observation plans may be updated accordingly. Data are analyzed in real time and made available to the relevant working group to guide further observation strategies based on preliminary results. Observations over multiple nights are thus possible and encouraged in case of a preliminary detection.

The system currently responds to alerts from \textit{Fermi}-GBM \cite{fermi_gbm}, \textit{Fermi}-LAT \cite{fermi_lat}, \textit{Swift}-BAT \cite{swift_bat)} and \textit{Swift}-XRT \cite{swift_xrt)} for GRBs, and from the LVK collaboration \cite{lvk_collab} for GWs, using tailored criteria for each event type. SVOM and Einstein Probe alerts currently require manual handling and are expected to be integrated in the near future. Same applies for the new Kafka-based format, to be adopted by GCN starting October 2025.

The full system is regularly tested using self-issued alerts that enter the system in the same way transient alerts do. This allows for reliable operations of all sub-components of the system.

\begin{figure}[htbp]
    \centering
    \includegraphics[width=0.7\textwidth]{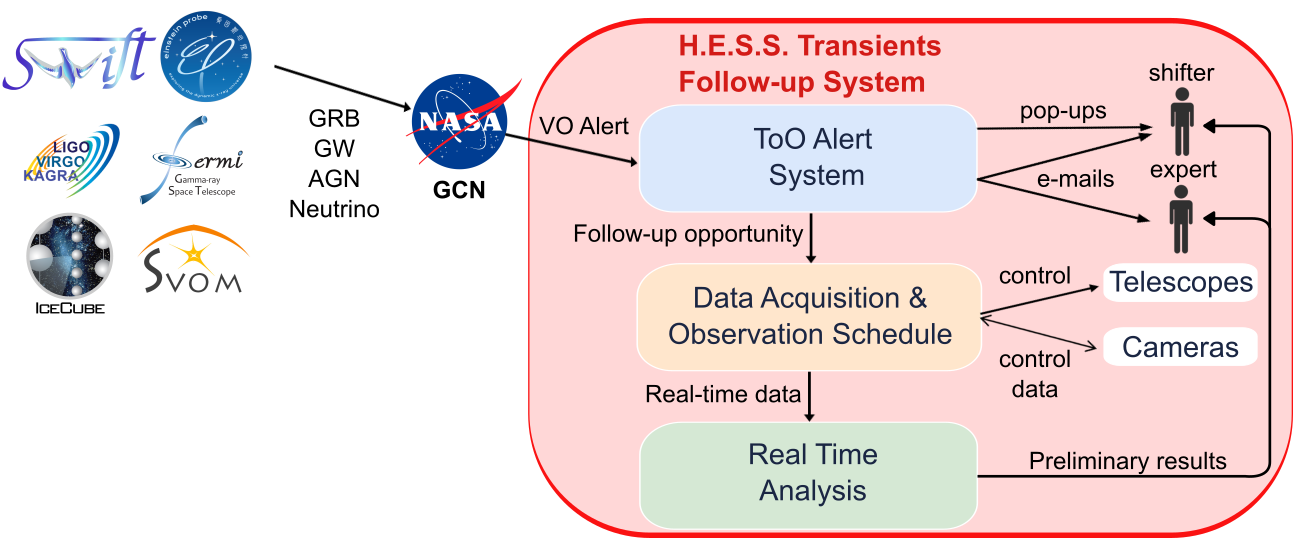}
    \caption{Schematic view of the H.E.S.S. transient follow-up system, illustrating the steps taken following the reception of an alert from multi-wavelength and multi-messenger observatories. Inspired from \cite{hess_gw)}. More details on each of the system blocks are available in the dedicated paper \cite{hess_follow_up_system}.}
    \label{fig:followup_software}
\end{figure}

\section{Follow-Up Statistics}
\subsection{GRBs Since 2020}

Statistics shown in this section are based on public alerts from the start of 2020 until June 2025, statistics and analyses of earlier alerts are presented in \cite{edna}. All H.E.S.S. GRB follow-ups since 2020 are documented in real time via a public GitHub page\footnote{https://grbhess.github.io/}.

\begin{figure}[htbp]
    \centering
    \begin{minipage}[b]{0.45\textwidth}
        \centering
        \includegraphics[width=\textwidth]{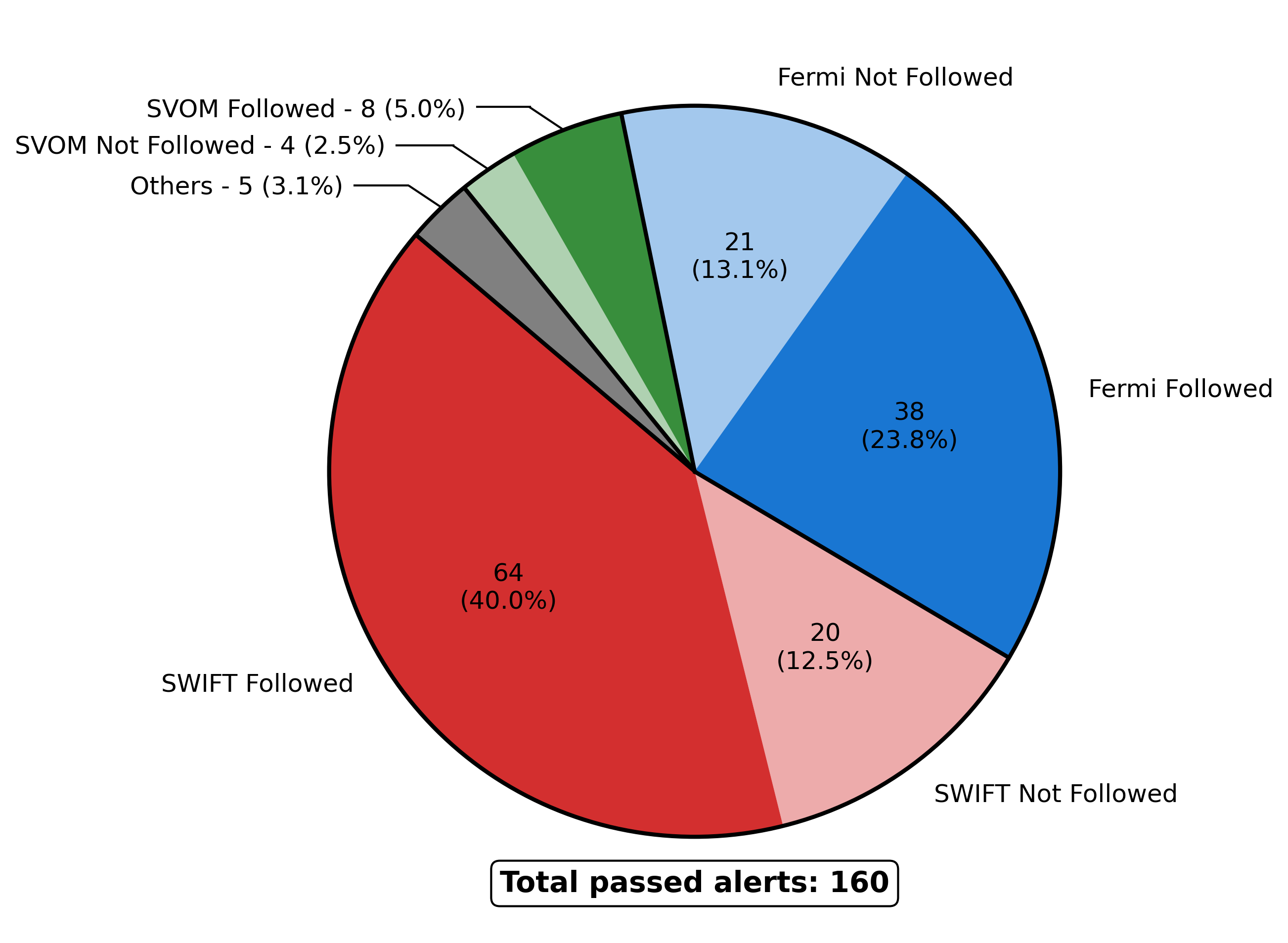}
    \end{minipage}
    \hspace{0.001\textwidth}
    \begin{minipage}[b]{0.45\textwidth}
        \centering
        \includegraphics[width=\textwidth]{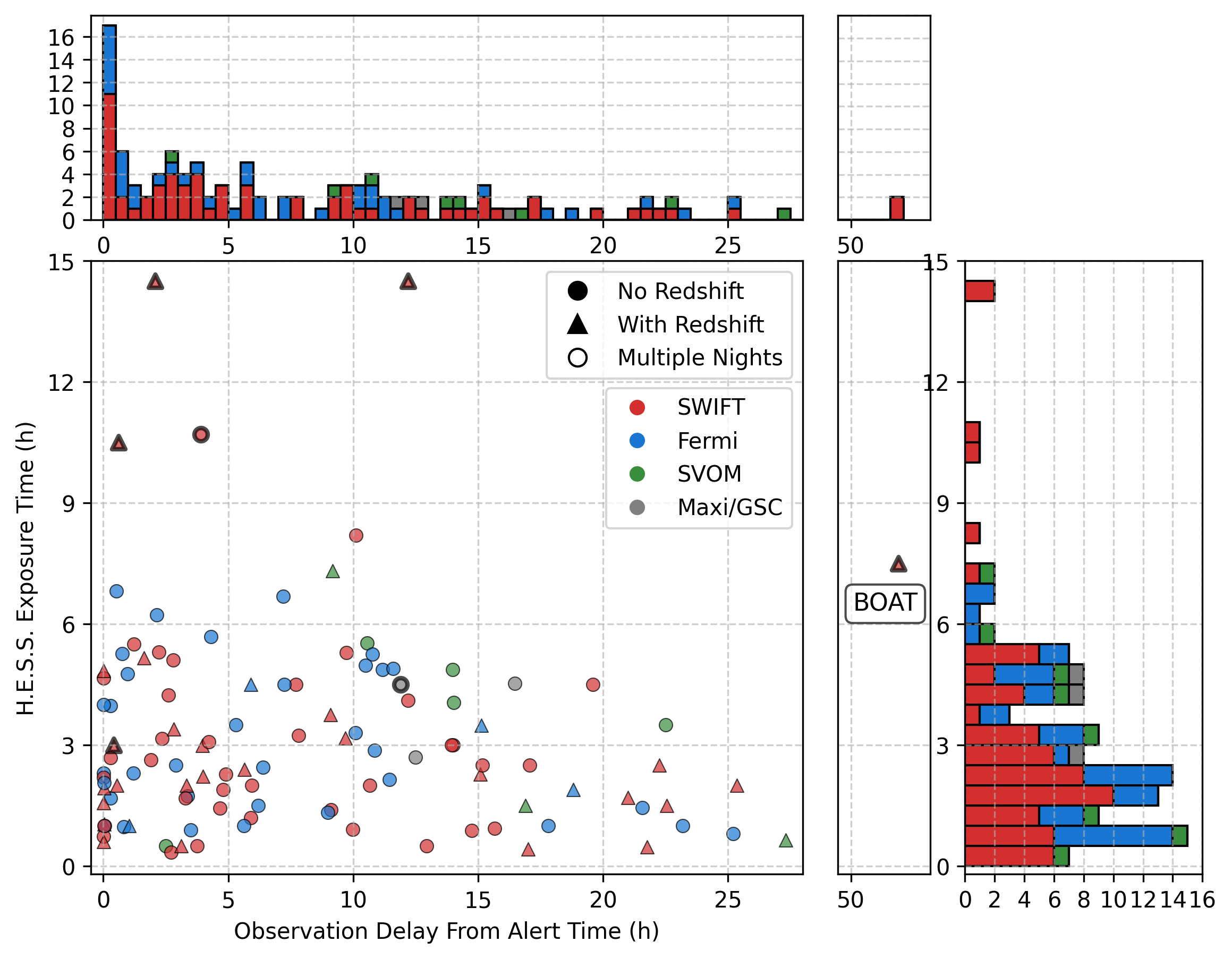}
    \end{minipage}
    \caption{Left: Pie chart describing all alerts that passed H.E.S.S. trigger criteria from 2020 to June 2025. From the 160 passed alerts, 113 were followed by the array whereas 47 were cancelled for various reasons (see text). Right: Scatter plot of H.E.S.S. exposure time versus GRB observation delay, with marginal histograms showing the distributions of both variables. Marker shapes indicate redshift information, and colors represent the alerting instrument. The first bin in the observation delay distribution corresponds to automatically triggered observations, consistent with the $\sim$15\% duty cycle of IACTs.}
    \label{fig:followup_statistics}
\end{figure}

Out of 1292 GRB alerts during this period, 160 (12.38\%) passed the H.E.S.S. trigger criteria. Forty-seven of these alerts were not followed due to various reasons (16 due to bad weather, 7 due to retracted alerts, 7 due to large positional uncertainty, 6 due to short observation windows, 5 identified as galactic transients, 2 with high redshift, 1 with high delay, and 3 canceled for other reasons). From the remaining 113 followed alerts, 35 have a listed redshift, measured either before or after the pointings. The distribution of followed/not-followed alerts per telescope is shown in a pie chart in the left side of Fig. \ref{fig:followup_statistics}, highlighting Fermi and Swift as the main trigger instruments for H.E.S.S. GRB follow-ups. All follow-ups are shown in Fig. \ref{fig:followup_statistics} (right), scattered in a plane showing the observation delay (difference between observation time and $t_{\text{event}}$, corresponding to the detection time by the trigger instrument) versus the telescope exposure time on the source. The delay takes into account the alert distribution time from the trigger observatories and the reaction time from the H.E.S.S. array; values in the plot are dominated by observability constraints.

\subsection{GWs During Run O4}

During the O4 run of the LVK collaboration, starting on May 2024, still ongoing as of June 2025 and expected to continue until November 2025, there has been a total of 203 non-retracted alerts. Out of these, 9 events were selected for ToO observations with H.E.S.S., including 2 bursts, 4 binary black hole (BBH) mergers, 2 likely neutron star-black hole (NSBH) mergers and 1 likely a binary neutron star (BNS) merger. Only the last three are expected to have a possible electromagnetic counterpart.

\section{Observation \& Analysis Examples}

This section presents two representative ToO follow-up cases to illustrate alert handling, observation strategy, and analysis procedures.

\begin{table}[h!]
\centering
\caption{Summary of the presented GRB and GW events follow-up parameters, gathered from public alerts.}
\label{tab:followup-summary}
\renewcommand{\arraystretch}{1.2} 
\scriptsize 
\begin{tabular}{@{}l|c|c@{}}
\toprule
& \textbf{GRB 240809A} & \textbf{S240422ed} \\
\midrule
\multicolumn{3}{@{}l}{\textbf{General information on the detection}} \\
\midrule
Trigger instrument & \textit{Swift}-BAT & LVK Collaboration \\
Distance           & z = 1.494 & 188 ± 43 Mpc  \\
Classification     & -- & NSBH (>99\%) \\
HasRemnant         & -- & 100\% \\
\midrule
\multicolumn{3}{@{}l}{\textbf{Final position}} \\
\midrule
Right ascension (J2000) [deg] & 237.55 & 123.31 \\
Declination (J2000) [deg]     & -2.32 & -26.07 \\
Position uncertainty    & 4.68 arcsec & 258 deg$^2$ \\
\midrule
\multicolumn{3}{@{}l}{\textbf{H.E.S.S. Observation Parameters}} \\
\midrule
Average zenith angle [deg]             & 39 & 30 \\
Delay of observation start w.r.t.\ $t_\mathrm{event}$ [h] & 9.25 & 115 \\
Exposure time [h]                          & 4 & 8 \\
\bottomrule
\end{tabular}
\end{table}

\subsection{GW Observation Example: S240422ed}

On April 22nd 2024, the LVK Collaboration detected an NSBH merger (classified as a likely NSBH merger with >99\% probability using the GstLAL pipeline \cite{gstlal}).

\begin{figure}[htbp]
    \centering
    \begin{minipage}[b]{0.3628\textwidth}
        \centering
        \includegraphics[width=\textwidth]{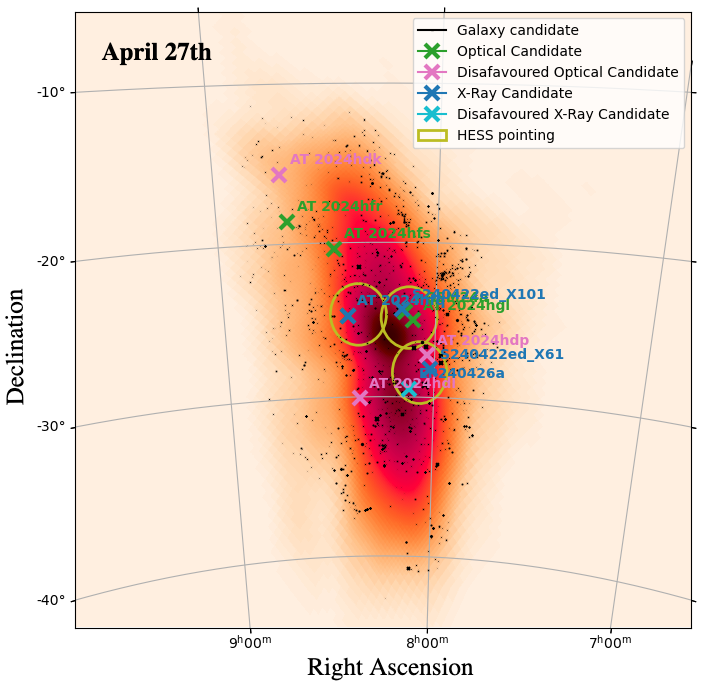}
    \end{minipage}
    \hspace{0.05\textwidth}
    \begin{minipage}[b]{0.39\textwidth}
        \centering
        \includegraphics[width=\textwidth]{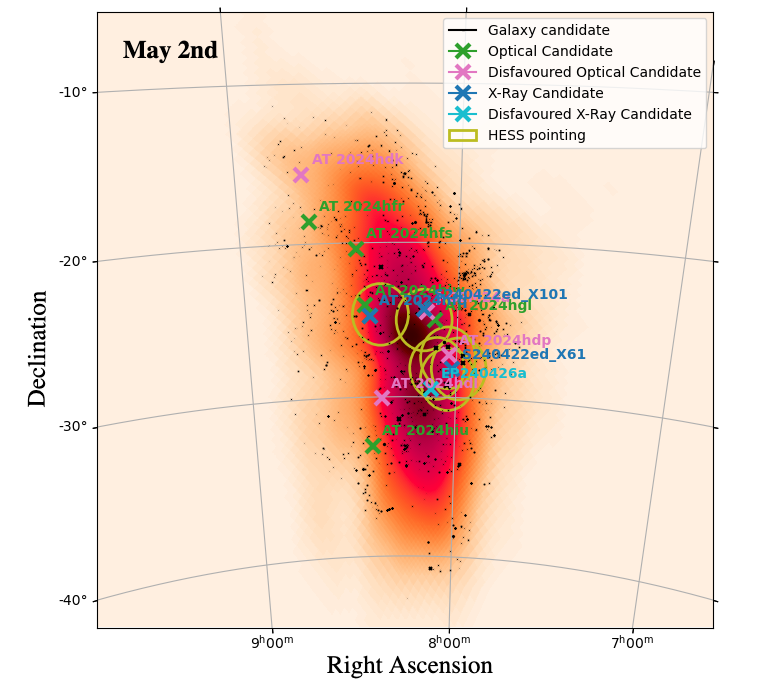}
    \end{minipage}
    \caption{Left: Multi-wavelength situation with possible counterparts of S240422ed on April 27th. Some of the possible counterparts were already ruled out at the moment. H.E.S.S. pointings tried to cover the most promising counterparts, especially X-ray sources. Right: Evolution of the multi-wavelength situation on May 2nd after five full nights of observations. Full H.E.S.S. pointings are represented, as well as new candidates and rejections from the intial ones.}
    \label{fig:tiling_GW}
\end{figure}

The source position became observable from the H.E.S.S. site only on April 27th. In the meantime, we closely followed multi-wavelength information provided in real time by the community through GCN circulars and notices, in order to maximize detection probability with an adapted follow-up strategy. For the first night of observations, 3 pointings were decided in order to cover the following possible counterparts: AT 2024hgl, S240422ed\_X101, J-GEM24a, S240422ed\_X61, AT 2024hdp, EP240426a, AT 2024hiw (Fig. \ref{fig:tiling_GW}, left). Following 2 nights of bad weather, the third night presented a hint of a hotspot in the preliminary analysis close to S240422ed\_X61, justifying observations for three more nights with wobble pointings around the hotspot. Fig. \ref{fig:tiling_GW} shows the evolution of the multi-wavelength panorama and how a follow-up is performed and adapted in real time in order to maximize the detection probability.

After off-line high-level analysis, a detection was ruled out. Moreover, a new classification performed by the LVK PyCBC pipeline \cite{pycbc} revealed a new likely terrestrial origin (95\% probability). Nevertheless, this event remains a representative example of a H.E.S.S. GW alert follow-up with a possible high-energy counterpart.

\subsection{GRB Analysis Example: GRB 240809A}

On August 9th 2024 \textit{Swift}-BAT detected a bright GRB with an optical counterpart, whose properties are summarized in table \ref{tab:followup-summary}. The event was also detected by Swift-XRT, which reported significant X-ray flux in the 0.3–10 keV energy range. 

Following the observations, the data were calibrated using two independent analysis pipelines. Primary analysis was performed with the Model++ framework \cite{hess_model_++}, and independently cross-checked with a separate calibration and reconstruction chain (HAP) \cite{hess_hap}. As with all ToO observations, low-level quality checks were conducted for each telescope and each observation run. Before any high-level analysis is performed, a dedicated unblinding procedure is applied to avoid bias.

\begin{figure}[htbp]
    \centering
    \begin{minipage}[b]{0.39\textwidth}
        \centering
        \includegraphics[width=\textwidth]{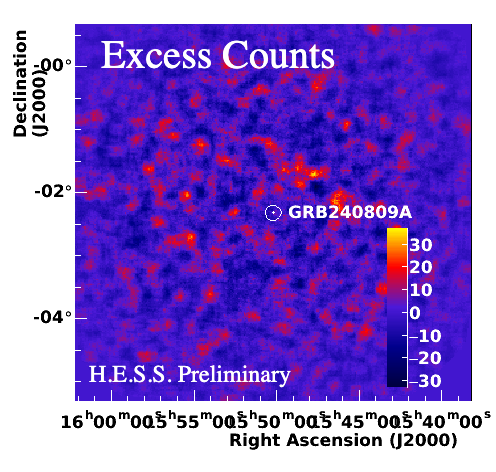}
    \end{minipage}
    \hspace{0.05\textwidth}
    \begin{minipage}[b]{0.39\textwidth}
        \centering
        \includegraphics[width=\textwidth]{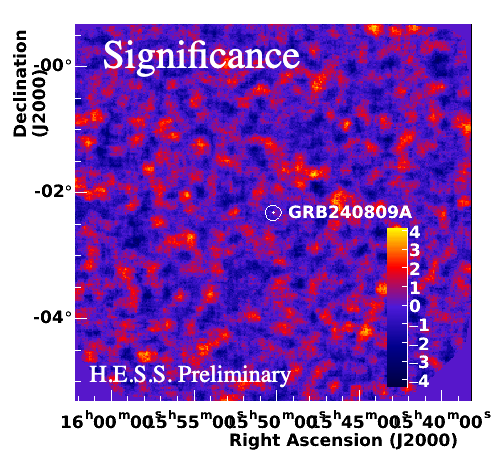}
    \end{minipage}
    \caption{Excess and significance maps derived from H.E.S.S. observations of the regions around GRB 240809A. The inner circles illustrate the size of the H.E.S.S. point spread function.}
    \label{fig:grb_sig_excess}
\end{figure}

Fig. ~\ref{fig:grb_sig_excess} shows the resulting excess counts and significance maps. No significant gamma-ray emission was detected within the Swift-XRT localization region. Given the precise localization, both differential and integral upper limits were computed at the GRB coordinates (95\% confidence). The analysis assumed a power law spectral model with index $\Gamma = 2$. The background was estimated using the reflected regions method \cite{bckg}, and the full analysis was performed using Gammapy v1.3 \cite{gammapy}.

\begin{figure}[htbp]
    \centering
    \begin{minipage}[b]{0.396\textwidth}
        \centering
        \includegraphics[width=\textwidth]{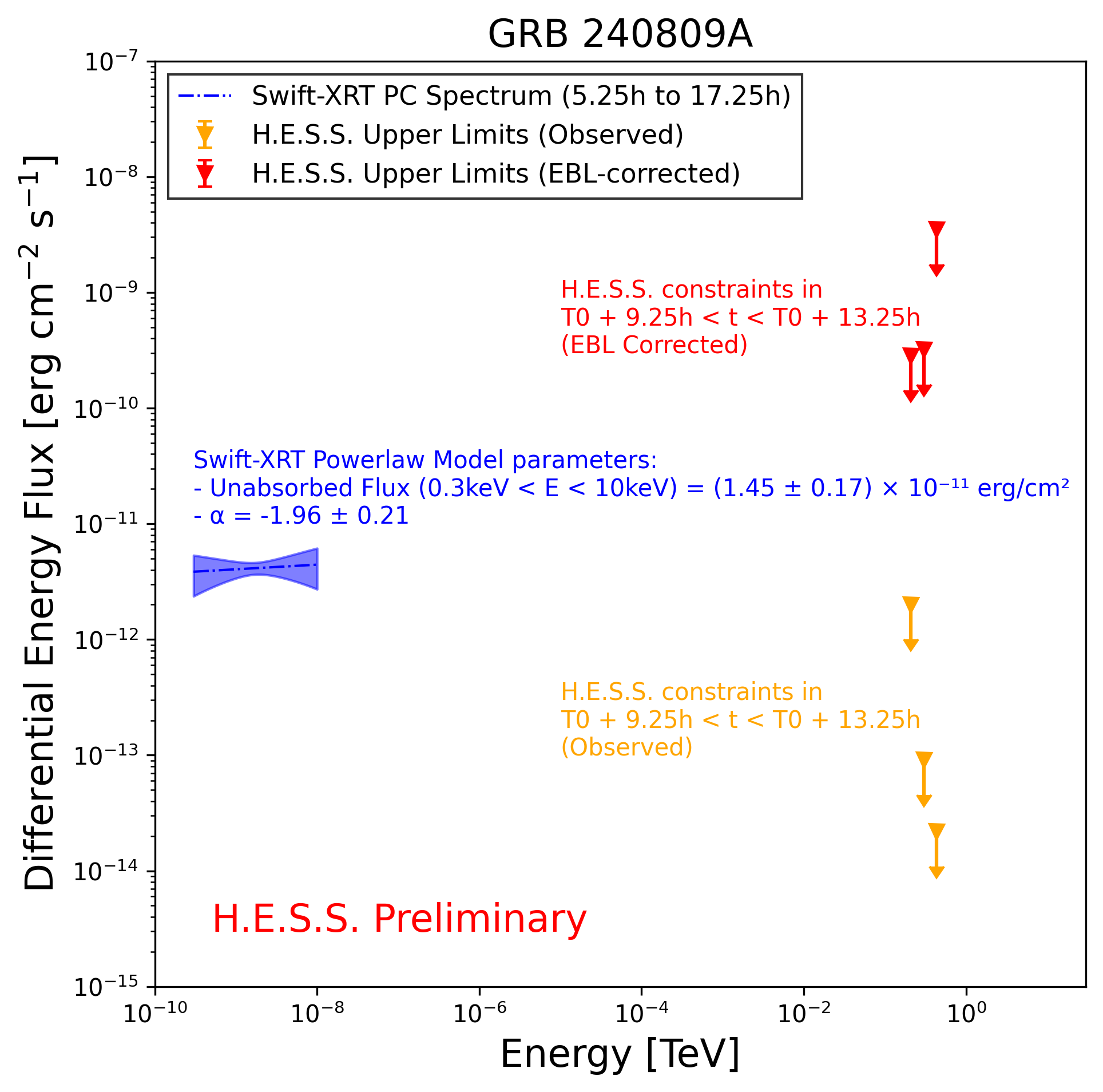}
    \end{minipage}
    \hspace{0.05\textwidth}
    \begin{minipage}[b]{0.39\textwidth}
        \centering
        \includegraphics[width=\textwidth]{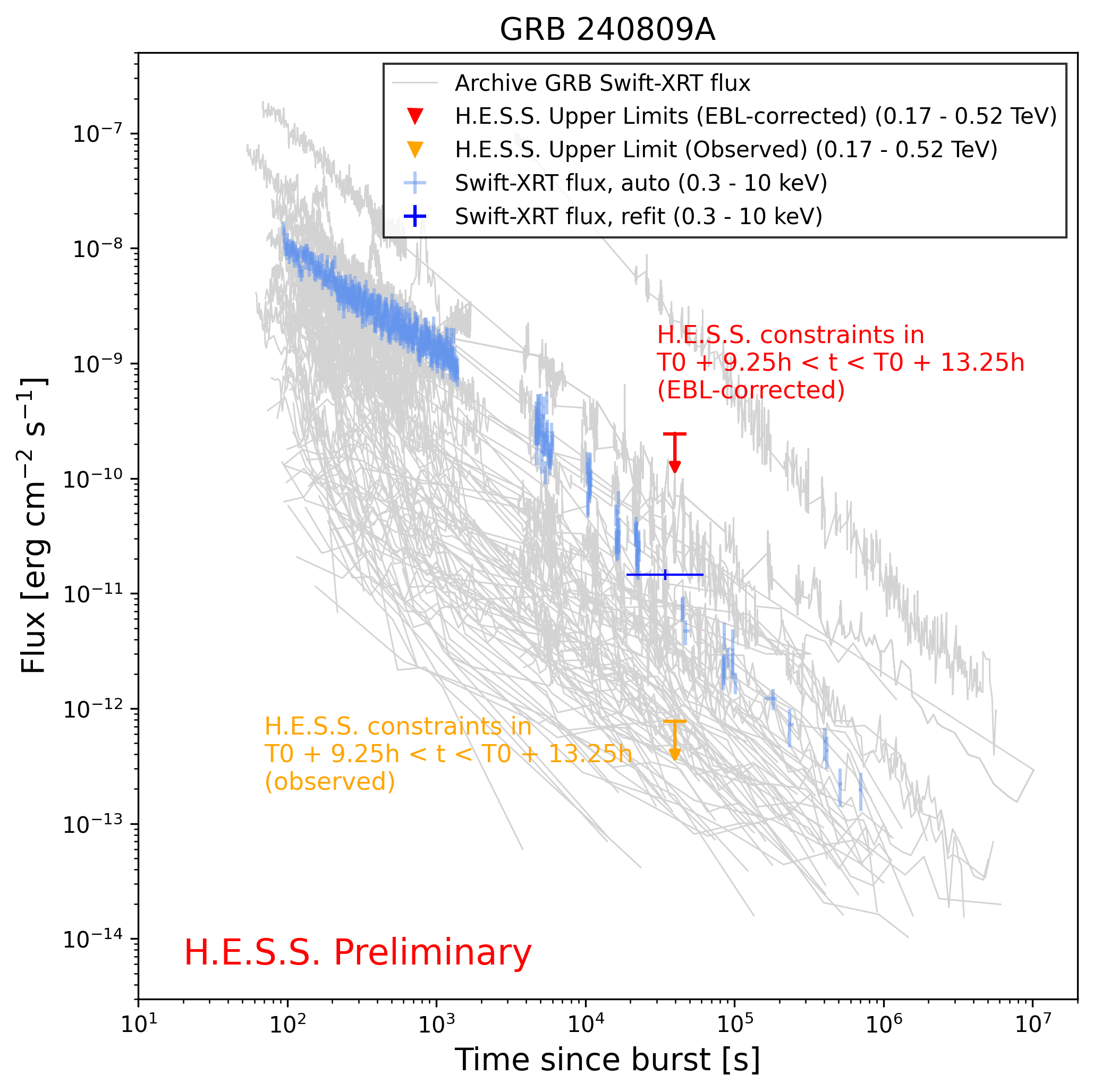}
    \end{minipage}
    \caption{Left: SED of GRB 240809A comprising Swift-XRT flux extracted using the time-sliced analysis tool from the Swift burst analyzer and H.E.S.S. upper limits computed on the same time range. Right: Light-curve with automatic Swift-XRT data extracted using swifttools and H.E.S.S. integral upper limits for an energy range of [0.17-0.52 TeV]. Multiple XRT observations around the H.E.S.S. observations $\pm$ 4h are then combined and refit (dark blue, 1$\sigma$ uncertainty).}
    \label{fig:grb_analysis}
\end{figure}

To place H.E.S.S. results in context, we compare the derived upper limits with \textit{Swift}-XRT data, extracted for the same observation time as H.E.S.S. using the time-sliced analysis tool from the Swift burst analyzer\footnote{https://www.swift.ac.uk/burst\_analyser/01247745/} \cite{swift_time_sliced}. Fig. ~\ref{fig:grb_analysis} shows both a spectral energy distribution (SED) and a light curve. The SED compares differential flux upper limits with XRT flux measurements over the same observation window. The light curve shows the evolution of the integral energy flux over time. The latter checks the evolution in time from gamma-ray emission with respect to X-ray emission, which have proven to have similar decay rates as the VHE-detected GRBs.

Both observed and EBL-corrected (using model \cite{dominguez})  upper limits are shown; only the latter should be directly compared with XRT data. The observed limits are included to illustrate the strong attenuation caused by EBL absorption for high-redshift sources. The energy range is constrained on the lower bound by H.E.S.S. sensitivity, reaching less than 200 GeV in the CT1-4 stereoscopic configuration, and by the EBL on the upper bound. 

Due to the source's high redshift, the derived limits are not expected to be strongly constraining. This event serves primarily as a representative example of a standard ToO analysis procedure within the H.E.S.S. transient follow-up program.

\section{Conclusion \& Outlook}

H.E.S.S. will continue to perform rapid follow-up observations of GRB and GW events, using information from a wide variety of trigger instruments and observatories. The recent launches of new X-ray missions such as SVOM and Einstein Probe, together with the optimized follow-up system developed by the H.E.S.S. collaboration, significantly enhance the prospects for a future very high-energy (VHE) detection.
Moreover, CT5 continues to be the world's largest Cherenkov Telescope. Combined with further improvements to the H.E.S.S. data analysis pipelines, it will allow us to significantly lower the H.E.S.S. energy threshold, not only for new observations but also for high potential archival observation. 
This new observational era will certainly give valuable insights on specific open questions in high-energy astrophysics, such as the connection between X- and gamma-ray emissions hinted by previous observations. Although these events are rare and require careful preparation, recent advances in follow-up strategies and data analysis techniques place H.E.S.S. in a strong position to detect the next VHE counterpart to a GRB or GW alert.

\section{Acknowledgements}

This work was carried out within the framework of the H.E.S.S. Collaboration. Acknowledgements can be found on the following page: https://hess.in2p3.fr/acknowledgements.

\begingroup
\setlength{\bibsep}{0pt plus 0.3ex}
\footnotesize
\bibliographystyle{unsrtnat}
\bibliography{refs/example.bib}
\nocite{*}
\endgroup

\end{document}